\documentclass[reprint,amsmath,amssymb,aps,prl,twocolumn,superscriptaddress,groupedaddress]{revtex4}
\usepackage{graphicx}
\usepackage{dcolumn}
\usepackage{bm}

\begin{document}

\title[Sample title]{Highly Mobile Carriers in a Candidate of Quasi-Two-Dimensional Topological Semimetal AuTe$_2$Br}

\author{Zeji Wang}
\affiliation{ICQM, School of Physics, Peking University, Beijing 100871, China}
\affiliation{Institute of Materials, China Academy of Engineering Physics, Jiangyou 621908, Sichuan, P. R. China}

\author{Shuyu Cheng}
\affiliation{ICQM, School of Physics, Peking University, Beijing 100871, China}

\author{Tay-Rong Chang}
\affiliation{Department of Physics, National Tsing Hua University, Hsinchu, Taiwan}

\author{Wenlong Ma}
 \affiliation{ICQM, School of Physics, Peking University, Beijing 100871, China}
 \author{Xitong Xu}
 \affiliation{ICQM, School of Physics, Peking University, Beijing 100871, China}
 \author{Huibin Zhou}
 \affiliation{ICQM, School of Physics, Peking University, Beijing 100871, China}
 \author{Guangqiang Wang}
 \affiliation{ICQM, School of Physics, Peking University, Beijing 100871, China}
 \author{Xin Gui}
 \affiliation{Department of Chemistry, Louisiana State University, Baton Rouge 70803, USA}
 \author{Haipeng Zhu}
 \affiliation{Wuhan National High Magnetic Field Center, Huazhong University of Science and Technology, Wuhan 430074, China}
 \author{Zhen Zhu}
 \affiliation{School of Physics and Astronomy, Shanghai Jiao Tong University, Shanghai, 200240, China}
 \author{Hao Zheng}
 \affiliation{School of Physics and Astronomy, Shanghai Jiao Tong University, Shanghai, 200240, China}
 \author{Jinfeng Jia}
 \affiliation{School of Physics and Astronomy, Shanghai Jiao Tong University, Shanghai, 200240, China}
 \author{Junfeng Wang}
 \affiliation{Wuhan National High Magnetic Field Center, Huazhong University of Science and Technology, Wuhan 430074, China}
 \author{Weiwei Xie}
 \affiliation{Department of Chemistry, Louisiana State University, Baton Rouge 70803, USA}
 \author{Shuang Jia}
 \thanks{gwljiashuang@pku.edu.cn}%
 \affiliation{ICQM, School of Physics, Peking University, Beijing 100871, China}
 \affiliation{CAS Center for Excellence in Topological Quantum Computation, University of Chinese Academy of Science, Beijing 100190, China}
 \affiliation{Kunshan Sunlaite New Energy Co., Ltd., Kunshan, 215300, China}

\date{\today}

\begin{abstract}
We report the crystal and electronic structures of a non-centrosymmetric quasi-two-dimensional (2D), candidate of topological semimetal AuTe$_2$Br. The Fermi surface of this layered compound consists of 2D-like, topological trivial electron and non-trivial hole pockets which host a Dirac cone along the $k_z$ direction. Our transport measurements on the single crystals show highly anisotropic, compensated low-density electrons and holes, both of which exhibit ultrahigh mobility at a level of  $10^5cm^2 V^{-1} s^{-1}$ at low temperature.The highly mobile, compensated carriers lead a non-saturated, parabolic magnetoresistance as large as $3\times10^5$ in single-crystalline AuTe$_2$Br in a magnetic field up to 58 T.
\end{abstract}

\maketitle

%

\section{\label{sec:level1}Introduction
}
The discoveries of Dirac and Weyl semimetal have motivated researchers' plethoric endeavor for sifting topological materials in crystallographic database.\cite{turner2013contemporary,weng2016topological,yan2017topological,armitage2018weyl} Topological semimetal (TSM) is characterized by topological non-trivial discrete band touching points or nodal lines near which the energy bands display linear crossing in the momentum space, forming Dirac and Weyl cones or nodal rings. The realistic quasi-electrons hosted on the cones can introduce many exotic electric properties including topological nontrivial surface state \cite{zheng2016atomic,zheng2016atomic2,zhu2018quasiparticle}, extremely high carrier mobility, large magnetoresistance (MR=$\frac{\rho_H-\rho_0}{\rho_0}$) and chiral anomaly \cite{hosur2013recent,jia2016weyl}. Till now, we have acknowledged various TSMs including Na$_3$Bi, Cd$_3$As$_2$, ZrTe$_5$, WTe$_2$ and TaAs family.\cite{xiong2015evidence,liang2015ultrahigh,li2016chiral,huang2015observation,zhang2016signatures}

Most of the previously reported TSMs show three-dimensional (3D) electronic structures, meaning their Fermi surfaces (FSs) mainly consist of 3D closed pockets in the momentum space. In general, time-reversal invariant 3D topological insulators (TIs) and TSMs are closely related to two-dimensional (2D) quantum spin Hall (QSH) insulators whose edge states are topologically protected.\cite{qi2011topological,hasan2010colloquium,weng2015quantum} Three dimensional TI and TSM can be built up by stacking the 2D QSH insulating layers along a certain crystal orientation while a QSH effect can be achieved in an exfoliated, monolayer TI and TSM. A well-studied example is the transition metal dichalcogenides (TMDs) whose monolayer atomic crystal can show QSH effect even at high temperature.\cite{wu2018observation,qian2014quantum} The bulk electronic structure of TSMs usually degenerates to 3D-like after stacking even though their crystal structures are still highly anisotropic.\cite{wu2017three,zhu2015quantum} This electronic structure evolution in dimension is reminiscent to the change from graphene to graphite.\cite{schneider2012using}

In this paper we introduce a layered candidate of TSM, AuTe$_2$Br whose electronic structure persists a 2D-like FS.
AuTe$_2$Br belongs to a family of halide AuTe$_2$X (X=Cl, Br, I) which crystalize in a quasi-2D structure consisting of halogen atoms sandwiched by AuTe$_2$ networks (Fig. 2a).\cite{rabenau1970telluride,haendler1974crystal,schmidt1977fermi,thesis}
By a comprehensive study of Shubonikov-de Haas (SdH) and de Haas-van Alphen (dHvA) quantum oscillations (QOs) in the single crystal, we found that the Fermi surface (FS) of AuTe$_2$Br mainly consists of separated hole and electron tubes with open orbits along the stacking direction.
Moreover, our transport measurements show that the electrons and holes are highly anisotropic while both exhibit ultrahigh mobility at a level of  $10^5cm^2 V^{-1} s^{-1}$ at low temperature.
As a result, single-crystalline AuTe$_2$Br manifests non-saturated, parabolic magnetoresistance as large as $3\times10^5$ in a magnetic field up to 58 T.
Such unique crystal and electronic structures naturally bridge the 3D TSM and 2D electron system, and also provide a platform for the fabrication of the devices based on its topological properties.

\section{EXPERIMENTAL METHOD}

We succeeded in synthesis of single crystalline AuTe$_2$Br, following the work of Rabenau et.al.\cite{rabenau1970telluride} 1.4 g gold powder (4N), 0.5 g tellurium powder (4N) and 0.5 mL liquid bromine were sealed in a fused silica ampoule ($\phi$ 10 mm$\times$L 10 cm, T 2.5 mm), and 65$\%$ of its capacity was filled with 9M hydrobromic acid.Then the ampoule was inserted into an autoclave with dry ice to apply balanced pressure outside at high temperature. The autoclave was heated to 350$^{\circ}$C and cooled down to 150$^{\circ}$C uniformly in 10 days.The as-grown samples are soft, silver flakes with a typical dimension of 2.5 mm $\times$2 mm $\times$ 0.2 mm, which is identical to previously reported.

For the single crystal XRD measurements, all of the single crystals from the samples were mounted on the tips of Kapton loop. The data were collected on a Bruker Apex II X-ray diffractometer with Mo radiation $K_1$($\lambda$= 0.71073\AA) and the measuring temperature is 100 K. Data were collected over a full sphere of reciprocal space with 0.5$^{\circ}$ scans in $\omega$ with an exposure time of 10 s per frame. The 2$\theta$ range extended from 4$^{\circ}$ to 75$^{\circ}$. The SMART software was used for data acquisition.Intensities were extracted and corrected for Lorentz and polarization effects with the SAINT program. Face-indexed numerical absorption corrections were accomplished with XPREP which is based on face-indexed absorption.\cite{sheldrick2000shelxtl} The twin unit cell was tested. With the SHELXTL package, the crystal structures were solved using direct methods and refined by full-matrix least-squares on $F^2$.\cite{sheldrick2008short} Crystallographic data are summarized in table I and II.

\begin{table*}
\caption{\label{tab:table1}Single crystal data of AuTe$_2$Br at 100 K. }
\begin{ruledtabular}
\begin{tabular}{p{5cm}<{\centering}|p{6.5cm}<{\centering}|p{6.5cm}<{\centering}}
Formula&AuTe$_2$Br
(centrosymmetric)&AuTe$_2$Br
(non-centrosymmetric)
\\
\hline
Temperature (K)&100 & 100\\
F.W. (g/mol) &532.08& 532.08 \\
Space group; $Z$&$Cmcm$ (No.63);4 & $Cmc2_1$ (No.36);4 \\
$a$(\AA) &4.018(3)& 4.018(3) \\
$b$(\AA) &12.284(7)& 12.284(7) \\
$c$(\AA) &8.885(5)& 8.885(5)\\
$V$(\AA$^3$)&438.5(5)&438.5(5)\\
Absorption Correction&Numerical&Numerical\\
Extinction Coefficient&0.00013(19)&0.00005(15)\\
$\theta$ range (deg)&3.317-30.428&3.317-30.428\\
No. reflections; $R_{int}$&1896; 0.1342&1896; 0.1328\\
No. independent reflections&390&595\\
No. parameters&16&27\\
$R_1$; $wR_2$ (all $I$)&0.0943; 0.1217&0.0929; 0.1056\\
Goodness of fit&1.201&1.131\\
Diffraction peak and hole (e$^-$/\AA$^3$)&4.606; -3.936&4.350; -4.355\\

\end{tabular}
\end{ruledtabular}
\end{table*}

\begin{table*}
\caption{\label{tab:table2}Atomic coordinates and equivalent isotropic displacement parameters of AuTe$_2$Br at 100 K. $U_{eq}$ is defined as one-third of the trace of the orthogonalized $U_{ij}$ tensor (\AA$^2$). }
\leftline{\bfseries Centrosymmetric\mdseries}
\begin{ruledtabular}
\begin{tabular}{ccccccc}
Atom&Wyckoff&Occupancy&x&y&z&$U_{eq}$\\
\hline
Au1 & 4c & 1 & 0 & 0.0817(1) & 1/4&0.0066(5)\\
Te2 & 8f & 1 & 1/2 & 0.1060(1) & 0.0542(2)&0.0062(5)\\
Br3 & 4c & 1 & 0 &-0.1652(3) & 1/4&0.0090(8)
\end{tabular}
\end{ruledtabular}

\leftline{\bfseries Noncentrosymmetric\mdseries}
\begin{ruledtabular}
\begin{tabular}{ccccccc}
Atom&Wyckoff&Occupancy&x&y&z&$U_{eq}$\\
\hline
Au1 & 4a & 1 & 0 & 0.4183(1) & 0.288(2)&0.0067(5)\\
Te2 & 4a & 1 & 1/2 & 0.3924(10) & 0.0920(0)&0.004(2)\\
Te3 & 4a & 1 & 1/2 &0.3954(9) & 0.4835(5)&0.008(2)\\
Br4 & 4a & 1 & 0 &0.6652(4) & 0.284(5)&0.0088(13)
\end{tabular}
\end{ruledtabular}

\end{table*}

The resistance and Hall resistance were measured via a standard four-point method, while the magnetic field dependent data in +H and -H were anti-symmetrized and symmetrized, respectively, in order to remove the non-symmetric part due to the contacts.The data in 9 T and 14 T were collected in a Quantum Design Physical Property Measurement System (PPMS-9) and Oxford TeslatronPT (14 T), respectively. Pulsed field measurement up to 58 T was performed at the Wuhan National High Magnetic Field Center. The torque measurement was performed by using a piezoresistive cantilever (SEIKO-PRC400) incorporated into the PPMS-9, on a crystal about 0.1 mm across and 0.01 mm thick of the same batch. A Wheatstone bridge circuit was adopted to detect a small change in the resistance.

\begin{figure}[htbp]
\includegraphics[clip, width=0.5\textwidth]{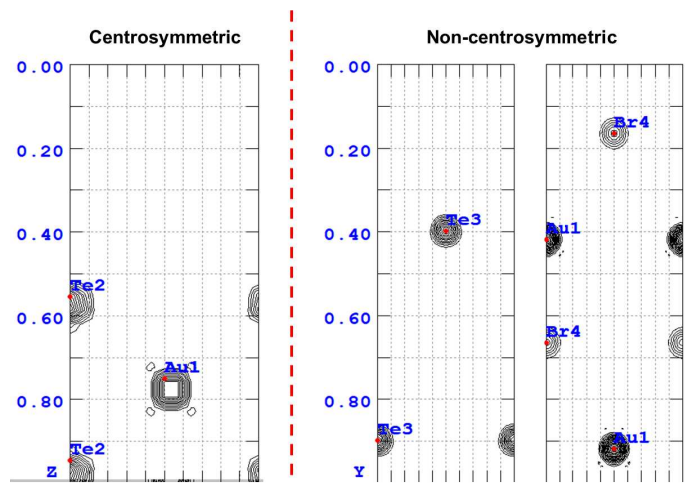}\\[5pt]
\caption{Observed electron distribution contours with centrosymmetric and non-centrosymmetric models.}
\end{figure}

\section{RESULTS AND DISCUSSION}

Our single crystal X-ray diffraction revealed that AuTe$_2$Br crystalizes in a non-centrosymmetric structure of the space group $Cmc2_1$ (a=4.0183\AA, b=12.284\AA, c=8.885\AA) which is different than previous report \cite{thesis}.
In order to confirm our result, we plot the electron distribution contours in centrosymmetric and non-centrosymmetric models (Fig. 1)
It is apparent that the observed electron distribution matches well with the non-centrosymmetric model.
Comparing with three sibling compounds, the 2D networks of AuTe$_2$ ripple in different shapes so that the Br and Cl ion has only one nearest neighboring Au atom whereas the Iodine ion has two equally nearest neighboring Au atoms in the upper and down layers, respectively. This structural difference makes the AuTe$_2$Cl and AuTe$_2$Br more 2D-like than AuTe$_2$I and therefore their crystals are soft and exfoliable, which promises the fabrication of the electric devices on few layers. Figure 2b shows a Scanning Tunnel Microscope (STM) image of an exfoliated crystal of AuTe$_2$Br which clearly unveils that the b plane consists of the unit cells of 4\AA$\times$8.9\AA, consistent with our XRD measurement.

\begin{figure}[htbp]
\includegraphics[clip, width=0.5\textwidth]{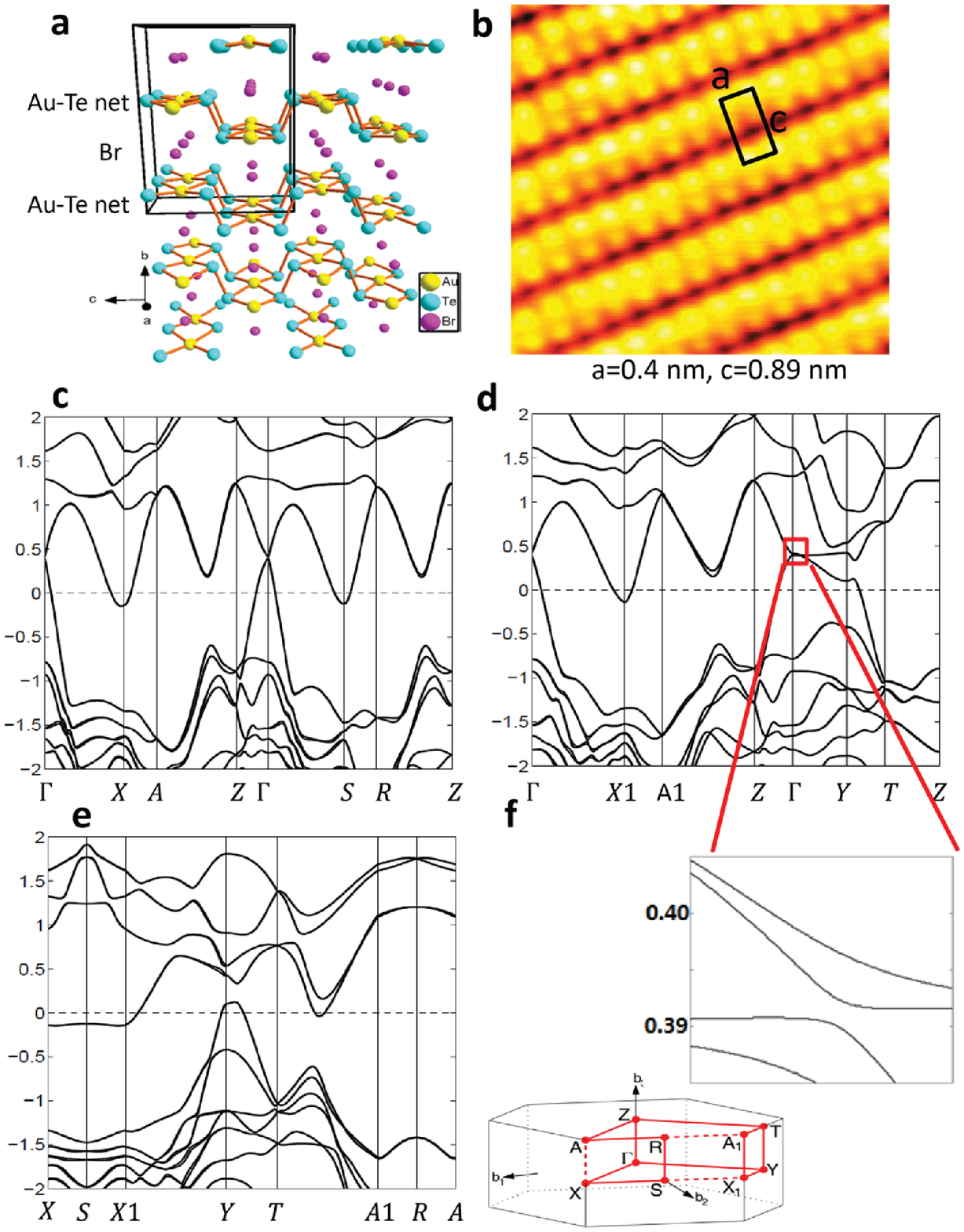}\\[5pt]
\caption{Crystal and Band Structures of AuTe$_2$Br. a: Crystal structure. b: An STM image of exfoliated surface. c-e: Bulk band structure in the represent of SOC. f: Zoom-in of the band dispersion near the $\Gamma$ point. Here we define the coordinates: $\bm{b_1}=\frac{2\pi}{\left|a\right|^2}\bm{a}+\frac{2\pi}{\left|b\right|^2}\bm{b}$; $\bm{b_2}=\frac{2\pi}{\left|b\right|^2}\bm{b}$; $\bm{b_3}=\frac{2\pi}{\left|c\right|^2}\bm{c}$ to describe the normal vectors for the BZ of AuTe$_2$Br. }
\end{figure}

Corresponding to its layered crystal structure, AuTe$_2$Br shows a quasi-2D electronic structure in the band dispersion.
Our band structure calculation reveals one hole-like and one electron-like pocket around the center and edge of the Brillouin Zone (BZ), respectively (Fig. 2c-e). The energy dispersion is much flatter in the ($b_1$, $b_2$) plane than along the vertical $b_3$ direction in the reciprocal space, indicating a quasi-2D nature of both electron and hole pockets. N. B. there exists a band-crossing Dirac cone in the hole pocket near the $\Gamma$ point above the Fermi energy ($E_F$) with no spin orbital coupling (SOC). When the SOC is turned on, the high symmetry line opens a very small gap near the crossing point (Fig. 2f) while the band structure near the $E_F$ remains intact.

A single crystal of AuTe$_2$Br shows an uncommonly good metallicity with a room-temperature in-plane resistivity $\rho\approx40 \mu\Omega cm$ (Fig. 3a). The crystal has a large residual resistance ratio (RRR=$\frac{\rho_{300K}}{\rho_{2K}}$ =576) and a small residual resistivity $\rho_{2K}=70 n\Omega cm$ in zero magnetic field. Such large RRR and small $\rho_{2K}$ are rarely observed in ternary compounds. As comparison, a crystal of Calaverite (AuTe$_2$) shows room-temperature resistivity $270 \mu\Omega cm$ and RRR=350.\cite{kitagawa2013pressure} The $\rho_{xx}$ is significantly enhanced under strong magnetic field below 50 K, but at high temperature it changes slightly. These profiles of temperature-dependent resistivity under different magnetic fields are similar to recently reported semimetals such as PtSn$_4$,\cite{mun2012magnetic} WTe$_2$,\cite{ali2014large} ZrSiS, \cite{singha2017large} and PtBi$_2$.\cite{gao2017extremely}

As observed in WTe$_2$, a perfectly compensated semimetal shows a quadratic magnetoresistance (MR) with no sign of saturation in strong field,\cite{zhu2015quantum,ali2014large} and that is what we observed for a piece of AuTe$_2$Br in a pulsed magnetic field up to 58 T at 4.2 and 10 K (Fig. 3b): the MR exceeds $3\times10^5$ at 4.2 K in 58 T with no sign of saturation. The apparent SdH QOs indicate that the carriers move along closed orbits when $\mathrm{H}\parallel \bm{b}$. In order to clarify the power law of the field dependence at higher temperatures, we measured the same sample'MR in a superconducting magnetic field up to 9 T. N. B. albeit the MR varies $10^{10}$ in 9~T at different temperatures in this double-logarithm scale plot, its power law remains intact as MR$\propto H^{1.9}$, very close to the quadratic behavior predicted in a Drude model for a compensated semimetal.

The large MR at 4.2 K for AuTe$_2$Br is comparable to the high-quality WTe$_2$ single crystal (RRR>1000),\cite{zhu2015quantum} which indicates its carriers also have a comparably high mobility. To estimate the mobility at low temperature, we used the relation between MR and magnetic field B=$\mu_0$H for a compensated semimetal:\cite{zhu2015quantum,ali2014large}$$\frac{\Delta\rho}{\rho_0}=MR=\mu_h\mu_eB^2\eqno(1)$$
A simple scalar mobility  $\overline{\mu}=\sqrt{\mu_h\mu_e}$ can be estimated as $1.1\times10^5  cm^2 V^{-1} s^{-1}$ at 4 K.

The densities and motilities of the compensated carriers are estimated by fitting the Hall resistivity $\rho_{yx}$ with respect to the field at different temperatures (Fig. 3c). The $\rho_{yx}$ is positively, linearly dependent on the field above 100 K, leading to  hole carrier concentration $n_h=1.45\times10^{20} cm^{-3}$ by a single band model. Below 50 K, the $\rho_{yx}$ shows non-linear dependence, concurrently with the large parabolic MR. A two-band model fitting below 50 K leads to $n_h  (n_e)$, denoting the carrier concentrations of holes (electrons), approximate to each other at a level of $10^{20}  cm^{-3}$. The $\mu_h$ and $\mu_e$, denoting the mobility of holes and electrons, are estimated as $3.5\times10^5  cm^2 V^{-1} s^{-1}$   and $0.7\times10^5  cm^2 V^{-1} s^{-1}$at 2 K, respectively.
These results are further confirmed by the two-band model fitting of the Hall conductivity ($\sigma _{xy}=\frac{\rho_{yx}}{\rho ^2_{xx}+\rho ^2_{yx}}$) in which we use the constrains zero-field resistivity ($\rho =e(\mu_h n_h+\mu_e n_e)$) and $MR=\mu_h\mu_eB^2$ (Fig. 3d). As shown in Fig. 3e, the fitting results are close to the fitting of $\rho_{yx}$ .

The existence of ultra-highly mobile carriers in AuTe$_2$Br can be confirmed by Drude model in which the conductivity in zero field ($\sigma=\rho^{-1}$) equals $e(\mu_h n_h+\mu_e n_e)$. Given $n_h\approx n_e\approx10^{20}  cm^{-3}$, we estimate that $\frac{\mu_h+\mu_e}{2}\approx5\times10^5  cm^2 V^{-1} s^{-1}$ at 2K. The three different methods give us a reliable estimation of highly mobile carriers in AuTe$_2$Br.

Given that the MR of AuTe$_2$Br is comparable to other high-mobility TSMs in magnitude, it is much more anisotropic than any TSMs reported before. Figure 3d shows extremely large anisotropic transversal angular-dependent MR whose profile is proportional to $\cos^2\theta$, indicating the dominant rule of the $\mu_0$H along the stacking $\bm{b}$ direction. If we assume the transversal conductivity $\sigma(\theta)\approx\rho^{-1}(\theta)$ equals the sum of conductivity of 2D and 3D electrons ($\sigma_{2D} (\theta)$ and $\sigma_{3D} (\theta)$), then$$\sigma(\theta)=\sigma_{2D}(\theta)+\sigma_{3D}(\theta)=\frac{\sigma_{2D}}{1+\overline{\mu}^2B^2\cos^2\theta}+\sigma_{3D}\eqno(2)$$

Assuming the transversal $\sigma_{3D}$ is independent on $\theta$, then the ratio of $\rho(\theta=0^{\circ})$ to $\rho(\theta=90^{\circ})$ gives the value of $\frac{\sigma_{2D}}{\sigma_{3D}}$  . This ratio is more than 100 under the magnetic fields higher than 3T and we infer that $\sigma_{2D}$ contributes more than $99\%$ of the total in-plane conductivity. As comparison, the ratios of $\frac{\rho(\theta=0^{\circ})}{\rho(\theta=90^{\circ})}$ is approximately 2 to 3 for previously reported quasi-2D TSMs AMnSb$_2$ and AMnBi$_2$ (A=alkaline earth) \cite{wang2011quantum,wang2012two,li2016electron,he2017quasi}and LaAgBi$_2$.\cite{wang2013quasi} Further investigation of angular-dependent MR is needed to elucidate the angular dependent oscillations and the sharp peak at $90^{\circ}$ and $270^{\circ}$ (see the caption of Fig. 3).

\begin{figure}[htbp]
\includegraphics[clip, width=0.5\textwidth]{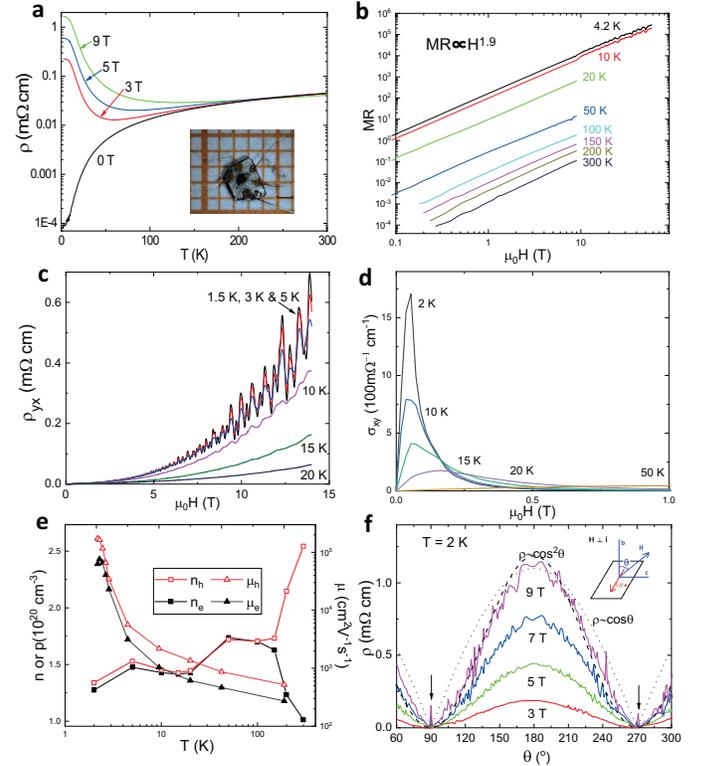}\\[5pt]
\caption{Electrical transport properties for AuTe$_2$Br. a: Temperature dependent resistivity in different magnetic fields. Inset: Photo of a single crystal. b: Parabolic MR in a bi-logarithmic coordinate at different temperatures. c: Hall resistivity with respect to field at low temperatures. d: Hall conductivity with respect to field at low temperatures. e: The carrier density and mobility at different temperatures fitted by two-band model. f: Angular-dependent transversal MR for $\mu_0H$ = 3, 5, 7 and 9 T. The inset illustrates the setup in which the current direction was kept perpendicular to the field. Note the angular-dependent MR manifests sharp peaks at $\theta=90^{\circ}$ and $270^{\circ}$ when the field is stronger than 5 T. This feature may be related to the coherent electron transport along small closed orbits on the sides of a corrugated cylindrical FS.\cite{singleton2002test,kikugawa2016interplanar}  }
\end{figure}

We now extract more information about the 2D and 3D carriers in AuTe$_2$Br from its field dependent QOs. As shown in Figure 4(a), the 2D hole pocket (named as $\beta$) is a gourd-like tube with peanut-shape maximum and minimum cross sections, while the 2D electron pocket (named as $\gamma$) is a corrugated, serpentine tube with nearly circular shape of cross section. The projected extremal orbits of $\beta$ and $\gamma$ and the small raisin-like electron ellipsoids (named as $\alpha$) along the $\bm{b}$ direction are shown in Fig. 4b.

The magnetic torque signal and electrical resistivity show significant QOs starting at 3 T at low temperatures when the field is along $\bm{b}$ direction (Fig. 4c). The oscillatory part of the torque and the MR with respect to reciprocal of the field ($\frac{1}{\mu_0H}$) is shown in Fig. 4d. Fast Fourier transform (FFT) discerns five base frequencies in SdH QOs which we named as $F_{\alpha}$, $F_{\beta1}$, $F_{\beta2}$, $F_{\gamma1}$ and $F_{\gamma2}$ and their multiplications (Fig. 4 e). Our result is apparently  different from previous report which shows four frequencies (52, 78, 105 and 173 T) in the SdH QOs \cite{SdHAuTe2Br}. Thanks to the simple topology of the FS, all the distinct frequencies can be identified completely corresponding to the extremal orbits of electron and hole pockets. The strong QOs of $F_{\beta1}$ and $F_{\beta2}$ stem from the peanut-shape cross sections of topological hole tube $\beta$, while the low frequency  $F_{\alpha}$ stems from the tiny 3D electron pocket $\alpha$. The two close weak QO frequencies $F_{\gamma1}$ and $F_{\gamma2}$ originate from the serpentine trivial electron tube $\gamma$ whose cross section along $\bm{b}$ is nearly uniform. The resistivity in pulsed fields up to 58 T reveals an identical FFT spectrum as that in low fields (Fig. 4 e). As comparison, the mapped Fermi surface in previous work seems to be inconsistent with the band structure calculation \cite{SdHAuTe2Br}.

\begin{table}
\caption{\label{tab:table3}The parameters of the electron and hole pockets from the analyses of SdH QOs in AuTe$_2$Br. The fitting results of dHvA QOs are shown in the brackets. Note: the calculated extremal cross section of $\gamma $ tube can not distinguish $F_{\gamma1}$ and $F_{\gamma2}$.}
\begin{ruledtabular}
\begin{tabular}{cccccc}
Cross Section&$F_{\alpha}$&$F_{\beta1}$&$F_{\beta2}$&$F_{\gamma1}$&$F_{\gamma2}$\\
\hline
F-measured (T) & 18.3 & 82.2 & 179.5 & 109.5 & 115.6\\
F-calculated (T) & 19 & 91 & 179 & \multicolumn{2}{c}{134} \\
$m^{\ast}(m_e)$ & 0.097 & 0.074(0.072) & 0.081(0.081) & 0.084 & 0.084\\
$k_F$ (\AA $^{-1}$) & 0.024 & 0.050 & 0.074 & 0.058 & 0.059\\
$v_F(10^5m/s)$ & 0.28 & 0.78 & 1.05 & 0.79 & 0.82\\
$E_F(eV)$ & 0.04 & \multicolumn{2}{c}{0.26 - 0.51} & \multicolumn{2}{c}{0.30 - 0.32}\\
$n/p(10^{19}cm^{-3})$ & 0.06 & \multicolumn{2}{c}{5.2} & \multicolumn{2}{c}{4.6}

\end{tabular}
\end{ruledtabular}
\end{table}

Identifying all the cross sections of the electron and hole pockets means we can address the carriers properties completely by analyzing their QOs. The parameters of the cyclotron masses ($m^{\ast}$)    for different frequencies are obtained by using Lifshitz-Kosevich (LK) formula:\cite{shoenberg2009magnetic}$$\frac{\Delta_{Amp}}{Amp}(T)=\frac{(2\pi^2k_BT/\beta)}{sinh(2\pi^2k_BT/\beta)}\eqno(3)$$
Where $\beta=e\hbar B/m^{\ast}$, $k_B$ is Boltzmann constant, $e$ is electron charge and $\hbar$ is Plank constant over $2\pi$. We fit the temperature dependent FFT amplitudes of the dHvA and SdH QOs for each frequency in Fig. 4f, and the results are shown in Table III. The cyclotron masses for each pocket are in a short range of $0.07 \sim 0.1 m_e$. Then we calculated the cross section $A_F=\frac{2\pi e}{\hbar}F$, the Fermi vector $k_F=\sqrt{\frac{A_F}{\pi}}$ and the Fermi velocity $v_F=\frac{\hbar k_F}{m^{\ast}}$  for each orbital. Note the maximum and minimum cross sections of $\beta$ tube are very different and therefore its parameters cannot be estimated precisely. The Fermi energy $E_F=m^{\ast}v_F^2$ is estimated for each pocket, and we found the values are consistent with the calculation (Fig. 2). The nearly identical parameters for  $F_{\gamma1}$  and  $F_{\gamma2}$ confirm their same origination from $\gamma$ tube.

In order to estimate the carrier densities of hole and electron tubes, we treat their FSs as cylinders with a height of $k_b=\frac{2\pi}{b}$, and then use the relationship $n=\frac{A_Fk_b}{4\pi^3}$ to get $n_{h-2D}=5.17\times10^{19}cm^{-3}$ and $n_{e-2D}=4.56\times10^{19}cm^{-3}$. This result is close to the two-band model fitting. The carrier density of the small 3D electron pocket was estimated as $n_{e-3D}=\frac{k_F^3}{3\pi^3}\approx0.06\times10^{19}cm^{-3}$, about $1\%$ of the 2D carriers. All above confirm AuTe$_2$Br as a compensated quasi-2D semimetal.

We can extract more information about the strongest QOs of  $F_{\beta1}$ which corresponds to the minimal cross section of the topological hole pocket $\beta$. The quantum life time $\tau_Q=\frac{\hbar}{2\pi k_BT_D}=6.3\times10^{-14}s$ is calculated by fitting the Dingle temperature $T_D=19.3 K$ in Fig. 4g at 2 K. Given  $\mu_h\approx3\times10^5  cm^2 V^{-1} s^{-1}$, the transport lifetime  $\tau_{tr}=\frac{\mu_hm^{\ast}}{e}$ is estimated as $1.3\times10^{-11} s$, leading to a ratio of $\frac{\tau_{tr}}{\tau_Q}=200$. The large ratio of transport lifetime and quantum lifetime is commonly observed in various TSMs.\cite{qi2011topological,zhang2017electron} It is well known that $\tau_{tr}$ measures backscattering processes that relax the current while $\tau_{Q}$ is sensitive to all processes that broaden the Landau subbands (LBs). The large ratio indicates that the small-angle scatterings are dominant while the backscattering is strongly protected at low temperature.
We notice that the SdH QO pattern is complicated because $F_{\beta2}$ is close to the double frequency of $F_{\beta1}$. Therefore it is difficult to extract the accurate phase information based our measurements.

\begin{figure}[htbp]
\includegraphics[clip, width=0.5\textwidth]{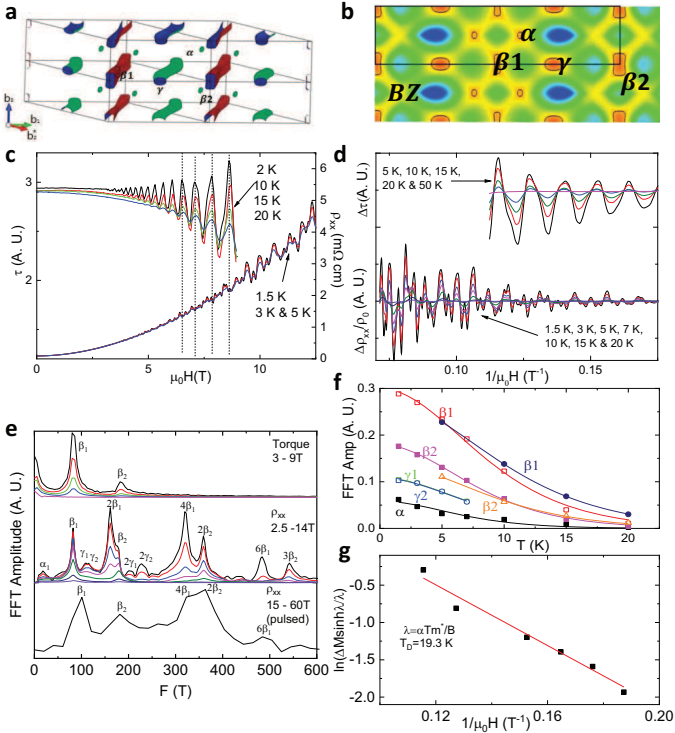}\\[5pt]
\caption{Quantum oscillations for AuTe$_2$Br. a: Calculated FS. Note here $\bm{b_2^\ast}=\frac{2\pi}{\left|a\right|^2}\bm{a}-\frac{2\pi}{\left|b\right|^2}\bm{b}$. b: The projected extremal cross sections of the orbits for $H\parallel b$. c: dHvA and SdH QOs with respect to the field at different temperatures. d: The oscillatory part with respect to $\frac{1}{\mu_0H}$ at different temperatures. e: The resultant FFT spectrum for dHvA QOs (upper); SdH QOs in low fields (middle) and in high fields (lower). f: Fitting of the cyclotron masses for different frequencies at different temperatures. g: Fitting of the Dingle temperature for $F_{\beta1}$ in dHvA QOs.}
\end{figure}


Figure 5 compares the SdH frequencies for the magnetic field along different directions with the calculation. The frequencies $F_{\beta1}$ and $F_{\beta2}$ approximately follow the $\frac{1}{\cos\theta}$  and $\frac{1}{\cos\varphi}$  relations at low tiled angles. The extremal cross section for the serpentine tube $\gamma$ changes in a complicated manner in high tilted angles in which the electrons can have several close extremal orbits. In experiment we do not observe any discernable frequencies from those orbitals when the tilted angle is higher than $60^{\circ}$. In contrast, the small electron ellipsoid changes in a different manner.

\begin{figure}[htbp]
\includegraphics[clip, width=0.5\textwidth]{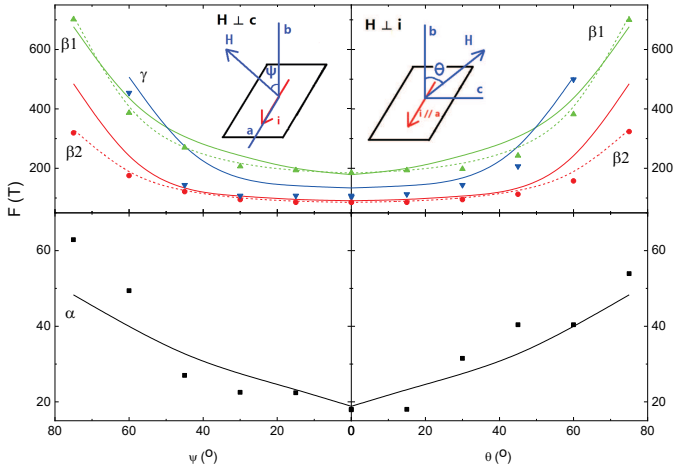}\\[5pt]
\caption{FS of AuTe$_2$Br. Change of the frequencies at different orientations. The dots are experimental results while the solid lines are calculated from the band structure. The dashed lines are $\frac{F_{\beta1,2}( 0^{\circ })}{\cos\theta}$ and $\frac{F_{\beta1,2}(0^{\circ })}{\cos\varphi}$. The tilted angels $\Psi$ and $\theta$ are defined in the insets.}
\end{figure}

The quasi-2D electronic structure of AuTe$_2$Br is closely related with its unique crystal structure. As shown in Fig. 2a, the AuTe$_2$ layer is distinct from the TMDs in which the transition metal has six coordinated chalcogen atoms. Here every Au atom has planar four-coordinated Te atoms which connect with other Te atoms in isolated Te-Te dimers (2.94\AA). The Te-Te dimers and Au atoms weave a ripple-like 2D network. This family of compounds have $18e$ in total if we account that the Au, Te and halogen atoms each contribute $13e$, $2e$ and $1e$, respectively. However we notice that the Te-Te dimer should share one pair of electrons and therefore it should follow $16e$ rule instead of 18. The $16e$ configuration is commonly observed in many transition metal square planar complexes containing Ir(I), Pd(II) or Au(III), in which the transition metal's $d_{x^2-y^2}$ orbital has significantly higher energy than the rest of the d orbitals.\cite{borgel2015transition} The molecular orbital (MO) $b_{1g}$ of $d_{x^2-y^2}$ orbital parentage is the lowest unoccupied for $16e$ count, which leads to a gap or pseudo-gap between the MOs of the rest d orbitals parentage.

 It is noteworthy that AuTe$_2$ can form different type of the layers in other quasi-2D layer compounds. For example, in the charge balanced $[\mathrm{Pb_2BiS_3}]^+[\mathrm{AuTe_2}]^-$ ,  the $[\mathrm{AuTe_2}]^-$ layer apparently obeys the $18e$ rule with no Te-Te dimers.\cite{fang2015two} Previous study revealed a 2D compensated semimetal which hosts highly mobile carriers as well. Future study of the devices based on the exfoliated two types of AuTe$_2$ layers will be interesting.

 In summary, we grew large single-crystalline AuTe$_2$Br and accomplished resistance and torque measurements in magnetic field. Our measurements reveal that it is a quasi-2D, compensated TSM hosting highly mobile electrons and holes. Its  unique crystal and electronic structure highlight a novel 2D topological electron system whose physical properties should be investigated in depth in the future.

\begin{acknowledgments}
The authors thank Chenglong Zhang, Gang Xu, Haizhou Lu, Jinglei Zhang and Qiang Gu for valuable discussions. This work was supported by the National Natural Science Foundation of China No. U1832214, No.11774007, the National Key R\&D Program of China (2018YFA0305601) and the Key Research Program of the Chinese Academy of Science (Grant No. XDPB08-1). S.J. was supported by Jiangsu Province Program for Entrepreneurial and Innovative Talents, Kunshan Program for Entrepreneurial and Innovative Talents and Natural Science Foundation of Jiangsu Province of China (BK2016). T.-R.C. were supported by the Ministry of Science and Technology under MOST Young Scholar Fellowship: the MOST Grant for the Columbus Program NO. 107-2636-M-006-004-, National Cheng Kung University, Taiwan, and National Center for Theoretical Sciences (NCTS), Taiwan. The work at LSU was supported by Beckman Young Investigator (BYI) program.

\end{acknowledgments}

\bibliographystyle{aipnum4-1}
\providecommand{\noopsort}[1]{}\providecommand{\singleletter}[1]{#1}%

\end{document}